\documentclass[aip,amsmath,amssymb,reprint]{revtex4-1}

\usepackage{graphicx}
\usepackage{dcolumn}
\usepackage{bm}

\usepackage{tabularx,ragged2e}
\newcolumntype{C}{>{\Centering\arraybackslash}X} 

\usepackage{multirow}

\usepackage{xcolor}

\begin{document}



\title{Transverse Oscillating Bubble Enhanced Laser-driven Betatron X-ray Radiation Generation}

\author{Rafal Rakowski}
\thanks{These two authors contributed equally}
\author{Ping Zhang}
\thanks{These two authors contributed equally}
\author{Kyle Jensen}
\author{Brendan Kettle}
\author{Tim Kawamoto}
\author{Sudeep Banerjee}
\author{Colton Fruhling}
\author{Grigory Golovin}
\author{Daniel Haden}
\author{Matthew S. Robinson}
\author{Donald Umstadter}
\author{B. A. Shadwick}
\author{Matthias Fuchs}
\email[]{mfuchs@unl.edu}
\affiliation{$^{1}$Department of Physics and Astronomy, University of Nebraska - Lincoln, Lincoln, Nebraska 68588, USA}
\date{\today}

\begin{abstract}
Ultrafast high-brightness X-ray pulses have proven invaluable for a broad range of research. Such pulses are typically generated via synchrotron emission from relativistic electron bunches using large-scale facilities. Recently, significantly more compact X-ray sources based on laser-wakefield accelerated (LWFA) electron beams have been demonstrated. In particular, laser-driven sources, where the radiation is generated by transverse oscillations of electrons within the plasma accelerator structure (so-called betatron oscillations) can generate highly-brilliant ultrashort X-ray pulses using a comparably simple setup. Here, we experimentally demonstrate a method to markedly enhance and control the parameters of LWFA-driven betatron X-ray emission. With our novel Transverse Oscillating Bubble Enhanced Betatron Radiation (TOBER) scheme, we show a significant increase in the number of generated photons by specifically manipulating the amplitude of the betatron oscillations. We realize this through an orchestrated evolution of the temporal laser pulse shape and the accelerating plasma structure. This leads to controlled off-axis injection of electrons that perform large-amplitude collective transverse betatron oscillations, resulting in increased radiation emission. Our concept holds the promise for a method to optimize the X-ray parameters for specific applications, such as time-resolved investigations with spatial and temporal atomic resolution or advanced high-resolution imaging modalities, and the generation of X-ray beams with even higher peak and average brightness.
\end{abstract}


\maketitle 

\section{Introduction}

Relativistic electron bunches can generate tunable high-brightness X-ray pulses via synchrotron emission. The generation of electron beams with relativistic energies typically requires kilometer-scale facilities. Recently, significantly more compact X-ray sources have been demonstrated using laser-wakefield accelerated (LWFA) electron beams \cite{Corde2013}. In particular, transverse (betatron) oscillations of electrons during acceleration in these plasma structures can generate ultrashort high-brightness X-ray pulses via synchrotron emission (so called betatron radiation) from a millimeter-scale interaction length \cite{Rousse2004,Kneip2010a}. Laser-driven betatron sources have matured to an extent suitable for use in applications in many research fields \cite{Fourmaux2011,Kneip2011,Cole2015,Mahieu2018,Hussein2019}. Their high brilliance enables advanced imaging methods and due to the ultrashort X-ray pulse duration and the intrinsic synchronization to the driving laser, these sources are ideal tools to investigate ultrafast dynamics \cite{TaPhuoc2007,Albert2016}. The comparably compact footprint and simple setup of these sources allows variable experimental geometries, including the potential for multiple time-synchronized X-ray pulses, at different wavelengths and at different incident angles for multi-probe diagnostics or single-shot tomography. This holds the promise of widespread availability for applications in diverse fields ranging from medical imaging and structural biology to chemistry, physics and materials science.

However, routine operation of these sources requires control over the source parameters, improvements in reproducibility and an increased laser-to-X-ray conversion efficiency. The latter is of particular importance for a potential reduction in the peak power of the driver laser, allowing for future sources with higher average flux at higher repetition-rate. Experiments have demonstrated that the X-ray spectrum can be enhanced through the direct interaction of the driver laser with the accelerating electrons \cite{Cipiccia2011}. Subtle laser characteristics, such as a pulse front tilt were used to enhance the X-ray radiation \cite{Yu2018}, manipulate the polarization \cite{Schnell2013} and spectrum \cite{Mangles2009}. An increase in photon flux has been shown by using lasers with a high pulse energy \cite{Kneip2008, Wang2013,Albert2017}. A higher conversion efficiency has been demonstrated by using clustered gas targets \cite{Chen2013} and the transverse deflection of the injected and accelerated electron beam through transversally offsetting the accelerating plasma structure by laser-electron-beam interactions \cite{Dong2018} or a modulated gas density profile \cite{Guo2019,Kozlova2020}.

Here we experimentally demonstrate a method to control and significantly enhance the radiation characteristics by manipulating the amplitude of the betatron oscillation during the electron injection process. Our approach is based on the evolution of the laser pulse into a highly asymmetric temporal shape (see Figure~\ref{fig4_simulations}a). This, in combination with the longitudinal dynamics of the accelerating plasma structure during a plasma-density downramp, leads to controlled injection of copious electrons from off-axis positions. These off-axis injected electrons perform large-amplitude collective transverse betatron oscillations during the subsequent propagation inside the accelerating structure similar to those in a permanent-magnet wiggler. We enable control over the source parameters by splitting the source into three sections: (i) laser pulse shaping, (ii) electron injection and (iii) a radiator section. These sections are implemented through a specifically tailored plasma density profile with a double-peaked ``M"-shaped structure (see inset Figure~\ref{fig1}). In the first density peak the laser pulse is modified into a highly asymmetric temporal shape through self-evolution, which leads to an LWFA where electrons are injected from off-axis positions \cite{Nerush2009}. The injection is controlled through a density downramp following the first peak, leading to a copious amount of electrons that are transversely injected with a longitudinally-correlated phase. The injected electrons perform coherent betatron oscillations during a density depression and in a second density peak. The density depression between the peaks and the density upramp of the second peak results in a nearly constant kinetic energy of the oscillating electrons (see section \ref{sec:PICSim}), leading to an increased X-ray brightness. While for certain combinations of plasma densities and laser intensities, transverse injection due to laser evolution also occurs in flat-top gas jets as was shown in simulations \cite{Ma2016}, our tailored plasma density profile enables control over these processes. Furthermore, it extends the length over which betatron radiation is emitted due to an increased laser depletion length of the lower integrated plasma density compared to a flat-top profile.

We were able to significantly enhance the generated number of X-ray photons when compared directly to targets with a flat-top density profile under identical laser conditions. In our concept, the coherent oscillations are initiated close to the gas jet entrance, unlike other methods where increased betatron oscillation amplitudes are achieved by manipulating the already injected and accelerated electron beam well into the gas jet \cite{Guo2019,Kozlova2020} or near the gas jet exit \cite{Dong2018}. This enables the control of multiple degrees of freedom and the potential for additional significant enhancement through emission from many more betatron oscillation periods, further increasing the X-ray brightness and laser-to-X-ray conversion efficiency. We demonstrate that the X-ray properties can be controlled through the manipulation of the betatron oscillation amplitudes by varying the plasma density profile.

Electrons in LWFAs operating in the self-injection ``bubble" regime \cite{Pukhov2002a} perform transverse betatron oscillations during acceleration due to off-axis injection and a linear transverse restoring force \cite{Kostyukov2009a,Kalmykov2009a}. The transverse acceleration causes betatron synchrotron emission into a narrow forward cone. The wiggler-like spectrum is characterized by the critical photon energy, which is given by \cite{Esarey:2002a}
\begin{equation}
E_{\mathrm{crit}} = \frac{3}{2}K\gamma^2 \frac{hc}{\lambda_\beta},
\end{equation}
with the electron relativistic factor $\gamma$, the betatron period $\lambda_\beta$, Planck's constant $h$ and the speed of light $c$. In the bubble regime, the electron deflection parameter $K$ is given by
\begin{equation}
K = r_\beta k_p\sqrt{\frac{\gamma}{2}}\approx \gamma \theta,
\end{equation}
with the betatron amplitude $r_\beta$ and the plasma wavenumber $k_p=\sqrt{n_e e^2/\epsilon_0 mc^2}$, where $n_e$ is the plasma density, $\epsilon_0$ the dielectric constant, $e$ the electron charge, $m$ the electron mass and $\theta$ the maximum angle of the electron velocity to the wiggler axis. For typical LWFA parameters (where $K \gg 1$) the number of emitted photons per oscillation period and electron scales as $N_{ph} \sim K$. Both, $N_{ph}$ and $E_{\mathrm{crit}}$ can be enhanced and controlled with a larger betatron amplitude. $N_{ph}$ can also be increased with a higher electron beam charge and a greater number of betatron oscillations, which we experimentally demonstrate here.

\section{Experimental Setup and Methods \label{exp_setup}}

\begin{figure}
\includegraphics[width=1\columnwidth]{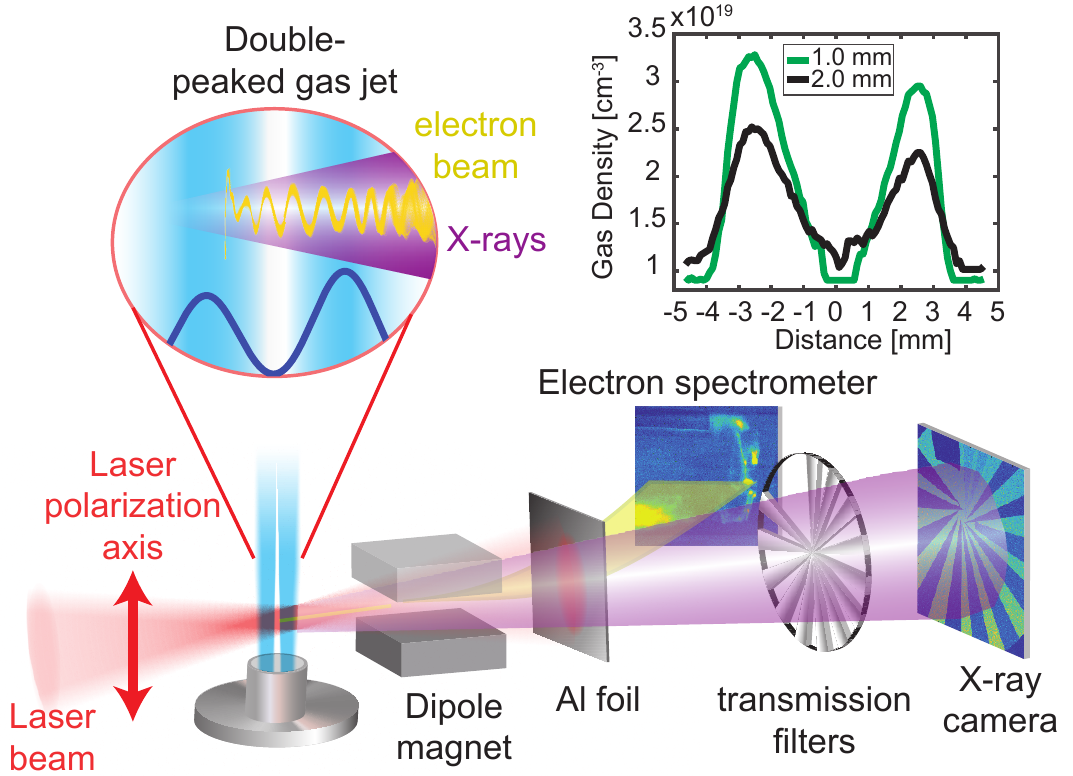}%
\caption{Schematic of experimental setup. A high-power laser (red) is focused into a double-peaked ``M'' shaped gas jet (blue). The laser evolution during the first density peak leads to off-axis electron injection during the following density downramp. Subsequent large-amplitude betatron oscillations (yellow) cause emission of intense X-ray radiation (purple). The laser pulse is filtered out by a thin Al foil. The electron bunch is deflected and characterized using a dipole magnet spectrometer. The X-rays are measured using an absorption-filter based spectrometer. The inset shows gas density measurements for a distance of $y$=1 mm (green) and $y$=2 mm (black) above the gas nozzle using a backing pressure of 250 psi, the betatron experiments were performed at lower densities.
\label{fig1}}
\end{figure}

In the experiment, we compare the betatron emission from different plasma density profiles, including an approximately 7-mm-long double-peaked ``M''-shaped profile (see inset Figure \ref{fig1}) to flat-top density profiles with lengths of 4 and 6 mm using the same laser parameters. The M-jet has an effective plasma length of approximately 4.4 mm consisting of two 2.2 mm wide peaks that are interrupted by a density depression of 2.3 mm. The betatron source is driven by laser pulses with an energy of $2.6 \,\, \mathrm J$ on target, a FWHM duration of 34 fs, focused to a 15 $\mu$m FWHM spot (see Figure \ref{fig1}). The laser-plasma interaction in the gas target causes the generation of relativistic electron bunches and intense X-ray pulses. The laser pulse is blocked by a thin Al filter, while the electron beam and X-rays above 2.7 keV are largely transmitted (see Supplemental Material). The electrons are deflected from the beam path using a dipole magnet (10.2 cm, 7.6 kG), which allows the simultaneous measurement of the electron and X-ray spectra in a single shot. The electron beam spectrum and divergence are measured using a phosphor screen located at a distance of 1.95 m from the jet. The X-ray spectrum is diagnosed by the transmission through a filter array consisting of Mylar, Al, V, Ti, Co, Fe, Zn, Cu, Zr, Mo, Pd, Ag, Sn and an Andor iKon-L SO X-ray CCD. The camera is shielded by a thin Al foil to minimize any optical background contamination. A Pb filter is used to determine any background. The filters were placed 1.9 m from the source and 0.75 m from the detector. The beamline filters and the quantum efficiency of the detector limited the spectral observation range to 2.7-30 keV (see Supplemental Material). The X-ray spectrum was deduced from least-square fitting of the transmitted signal through each filter, taking into account the transmission through additional beamline filters and the quantum efficiency of the detector (for detailed description see Methods section below). To account for the broad electron energy spread and the changing energy of the emitting electrons during the acceleration process, the spectral intensity was constrained to a sum of synchrotron spectra given by \cite{Elleaume}
\begin{equation}
	\frac{d^2 \phi}{d\Omega dE/E} = \sum_i^N A_i \left( \frac{E}{E_{\mathrm{crit},i}} \right)^2 F^2_{2/3}[E/(2E_{\mathrm{crit},i})],
	\label{eqn:synchrotron_spectra} 
\end{equation}
where $\Omega$ is the solid angle, $E$ the photon energy and $F$ is the modified Bessel function of the second kind. The amplitudes $A_i$ and critical energies $E_{\mathrm{crit},i}$ are fit parameters. We limit our model to two sub-spectra ($N$=2) to avoid over-fitting the data while showing a significant improvement over a single spectrum. A large set of Monte-Carlo (MC) simulations was used in a bootstrap method \cite{Press2007} to determine the confidence intervals via a statistical approach (see Methods).

\section{Experimental Results}

\subsection{X-ray Spectra}
\label{sec:X-ray Spectra}
\begin{figure}
\includegraphics[width=1\columnwidth]{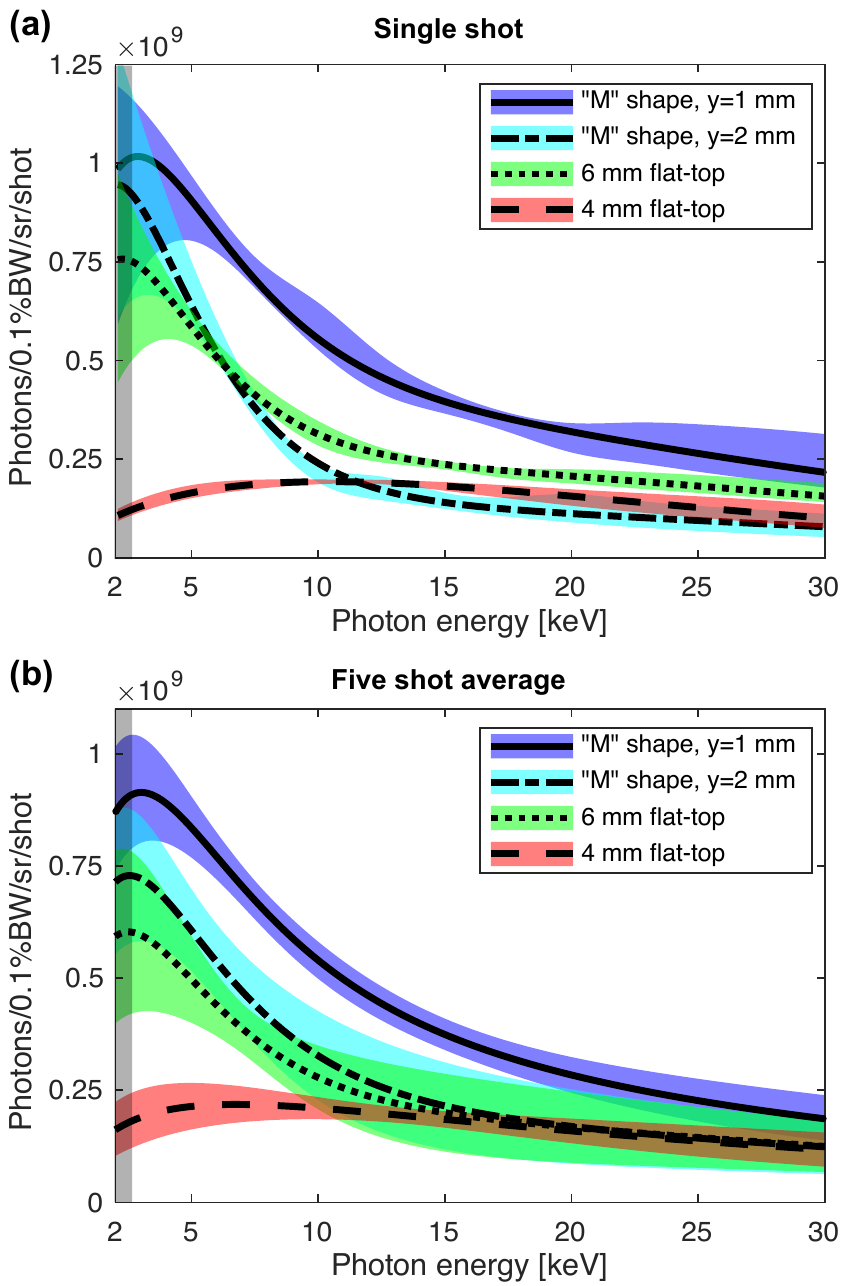}%
\caption{Measured X-ray Spectra. (a) shows  spectral intensity extracted from the filter transmission for a flat-top 4 mm jet (red, dashed), a flat-top 6 mm jet (green, dotted) and the ``M''-shaped jet for an interaction distance above the nozzle of $y_1=1 \,\,\mathrm{mm}$ (blue, solid) and $y_2=2 \,\,\mathrm{mm}$ (cyan, dash-dotted) for single shots. (b) shows a five-shot average. The fit parameters for the five-shot average are given in Table \ref{tab:table1}. The colored shades represent the confidence bands obtained through a statistical analysis of a large set of Monte-Carlo (MC) simulations. The confidence bands for the single shots indicate a point-wise one-sigma band of bootstrapped datasets. The confidence band for the averaged shots are obtained by uncertainty propagation using the individual shot uncertainties (see Methods section). The spectral observation range is limited to 2.7-30 keV (indicated through vertical gray shades) due to filter transmission and the quantum efficiency of the CCD camera. \label{fig2_Xrays}}
\end{figure}

\begin{table*}
\setlength\extrarowheight{2pt}
\caption{\label{tab:table1} Gas jet comparison. The angular flux density is averaged over 5 shots for each jet. The total number of photons is estimated from the angular flux density and the beam divergence. As the X-ray beam divergence exceeds the detector solid angle, the horizontal X-ray beam divergence is inferred from the electron beam divergence and the vertical divergence from the angular acceptance of the CCD detector (11 mrad). The averaged critical energies $E_{\mathrm{crit,}i}$ and amplitudes $A_i$ for each jet are obtained by fitting an $N=2$ sum of synchrotron spectra (equation \ref{eqn:synchrotron_spectra}) to the 5-shot averaged spectra. The uncertainties are obtained by fitting a similar term to the 5-shot confidence bands.}
\begin{ruledtabular}
\begin{tabular}{c|c|c|c|c|c }
Gas jet&Angular flux density&Beam divergence&Estimated total number of photons &\multicolumn{2}{c}{Synchrotron spectra fits}\\ \cline{5-6}
 & [photons/sr/shot] & [mrad$^2$] & [photons/shot] & $E_{\mathrm{crit},1}$ [keV]  & $E_{\mathrm{crit},2}$ [keV]  \\
 &  (2.7 - 30 keV)  &   & (2.7 - 30 keV) &  (Amplitude $A_1$) & (Amplitude $A_2$) \\ \hline
\multirow{2}*{\shortstack{4 mm \\ flat-top}} & \multirow{2}*{$0.5 \, \substack{+0.1 \\ -0.1} \times 10^{12}$} & \multirow{2}*{$24 \times 11$} & \multirow{2}*{$1 \, \substack{+0.2 \\ -0.2}\times 10^8$} & $4.4 \, \substack{+0.03 \\ -0.6}$  & $16.0 \, \substack{+5.0 \\ -5.1}$ \\
 & & & & $(0.6 \, \substack{+0.4 \\ -0.01}\times 10^8)$ & $(1 \, \substack{+0.2 \\ -0.01}\times 10^8)$\\[2pt] \hline

\multirow{2}*{\shortstack{6 mm \\ flat-top}} & \multirow{2}*{$0.8 \, \substack{+0.2 \\ -0.2} \times 10^{12}$} & \multirow{2}*{$10 \times 11$} & \multirow{2}*{$1 \, \substack{+0.2 \\ -0.2}\times 10^8$} & $2.7 \, \substack{+0.9 \\ -0.4}$ &  $15.6 \, \substack{+2.5 \\ -1.0}$\\
 & & & & $(3.3 \, \substack{+0.9 \\ -0.7}\times 10^8)$& $(1.2 \, \substack{+0.7 \\ -0.7} \times 10^8)$ \\[2pt] \hline
 
\multirow{2}*{\shortstack{M-jet \\ $y$=1 mm, 65 psi}} & \multirow{2}*{$1.5 \, \substack{+0.2 \\ -0.1} \times 10^{12}$} & \multirow{2}*{$ > 37 \times 11$} & \multirow{2}*{$ >6 \, \substack{+0.6 \\ -0.6}\times 10^8$} & $3.1\, \substack{+0.3 \\ -0.3}$&  $12.7\, \substack{+0.9 \\ -1.3}$\\
 & & & & $(4.6 \, \substack{+0.8 \\ -0.7}\times 10^8)$& $(2.2 \, \substack{+0.4 \\ -0.3}\times 10^8)$ \\[2pt] \hline

\multirow{2}*{\shortstack{M-jet \\ $y$=2 mm, 90 psi}}& \multirow{2}*{$1.0 \, \substack{+0.2 \\ -0.2} \times 10^{12}$} & \multirow{2}*{$ > 37 \times 11$} & \multirow{2}*{$ >4 \, \substack{+0.9 \\ -0.9}\times 10^8$} & $2.8\, \substack{+0.5 \\ -0.4}$ &  $15.0\, \substack{+2.4 \\ -2.3}$\\
 & & & & $(4.1 \, \substack{+0.5 \\ -0.5}\times 10^8)$& $(1.2 \, \substack{+0.9 \\ -0.8} \times 10^8)$ \\[2pt]
\end{tabular}
\end{ruledtabular}
\end{table*}

For each jet the X-ray flux was optimized by adjusting the plasma density and the jet position relative to the laser focal plane. The brightest X-ray pulses with the 4 and 6 mm flat-top jets were generated for a plasma density of $5 \times 10^{18} \,\, \mathrm{cm}^{-3}$ and with the M-jet for $1 \times 10^{19} \,\, \mathrm{cm}^{-3}$. The observed spectral intensity for the ``M''-shaped profile shows a clear enhancement in comparison to the other jets (see Figure \ref{fig2_Xrays}). Since the X-ray divergence exceeds our detector solid angle of $11 \,\, \mathrm{mrad}$, we estimate the total number of X-ray photons in a shot using the electron beam divergence. The total X-ray beam divergence is given by the convolution of the single-electron emission and the maximum electron beam deflection angle inside the bubble. For a wiggler or betatron source, the X-ray divergence is dominated by the electron beam deflection \cite{Elleaume}, which in the LWFA bubble regime is equivalent to the electron beam divergence after termination of the plasma, i.e. $\theta_{e,\mathrm{beam}} \approx K/\gamma$ $^[$\cite{Esarey:2002a}$^]$. The X-ray divergence in this case is approximately given by $\theta_{X,\mathrm{beam}}\approx \theta_{e,\mathrm{beam}} \approx K/\gamma$. The electron beam generated by the 6 mm jet has a horizontal divergence of 10 mrad, whereas that from the M-jets are transversally clipped by the spectrometer dipole magnet with an aperture of 37 mrad (see Figure 3). Using these values in the laser polarization dimension and the solid angle of the detector in the other dimension as conservative lower limits, we estimate an increase in the number of X-rays generated by the M- jet to be nearly one order of magnitude compared to the 4 mm and the 6 mm round jets (see Table \ref{tab:table1}). The sharp edges of the filter shadows indicate that the signal originates from a small source point and is not due to Bremsstrahlung background from scattering of the electron beam on the chamber walls.
 
We can estimate the critical photon energy from the electron beam parameters using equations (1) and (2) and the betatron period, which in the bubble regime is given by $\lambda_\beta = \sqrt{2\gamma} 2\pi/k_p$ $^[$\cite{Esarey:2002a}$^]$. Using the electron beam divergence to estimate $K$, we obtain for the M-jet a maximum $E_\mathrm{crit} = 186 \,\, \mathrm{keV}$ ($\theta_e$= 37 mrad, $\gamma$= 1100, $n_e = 1 \times 10^{19} \,\ \mathrm{cm^{-3}}$), for the 6 mm jet a maximum $E_\mathrm{crit} = 5 \,\, \mathrm{keV}$ ($\theta_e$= 10 mrad, $\gamma$= 490, $n_e = 5 \times 10^{18} \,\,\mathrm{cm^{-3}}$) and for the 4 mm jet a maximum $E_\mathrm{crit} = 12 \,\, \mathrm{keV}$ ($\theta_e$= 24 mrad, $\gamma$= 500, $n_e = 5 \times 10^{18} \, \,\mathrm{cm^{-3}}$). For the 4 mm jet, this estimate agrees reasonably well with the experimentally determined value. In case of the 6 mm jet, the spectrum extends to higher photon energies well beyond the estimated critical energy. This is likely due to the X-ray emission from electrons with higher energies at some point during the interaction, which occurs for a jet that is longer than the laser depletion length and the dephasing length (see section \ref{sec:Discussion}).  The experimentally measured values for the critical energies of the M-jet are significantly lower, which is a result of the limited sensitivity of our spectrometer of 2.7-30 keV. Therefore, we expect a substantial fraction of the generated betatron photons to be outside our spectral detection limits, particularly in case of the M-jet. Considering the photons outside our angular and spectral acceptance, we speculate the actual enhancement in photon number to be even significantly larger for the M-jet to compared to the flat-top profiles.

We also demonstrate that we can control the X-ray parameters by modifying the plasma density profile. To this end we increase the distance between the laser-plasma interaction point and the gas nozzle. As can be seen from the density measurement in Figure \ref{fig1}, the density gradient and the peak-to-valley contrast can be modified by changing the distance between the nozzle and the interaction point. We compare our measurements with a distance of $y_1 = 1 \,\, \mathrm{mm}$ to $y_2 = 2 \,\, \mathrm{mm}$ at similar plasma densities by compensating the gas jet backing pressure. The larger distance leads to a shift of the spectrum towards lower photon energies and an overall lower number of photons (see Figure \ref{fig2_Xrays} and Table \ref{tab:table1}).

\subsection{Electron Spectra}
\begin{figure}
\includegraphics[width=1\columnwidth]{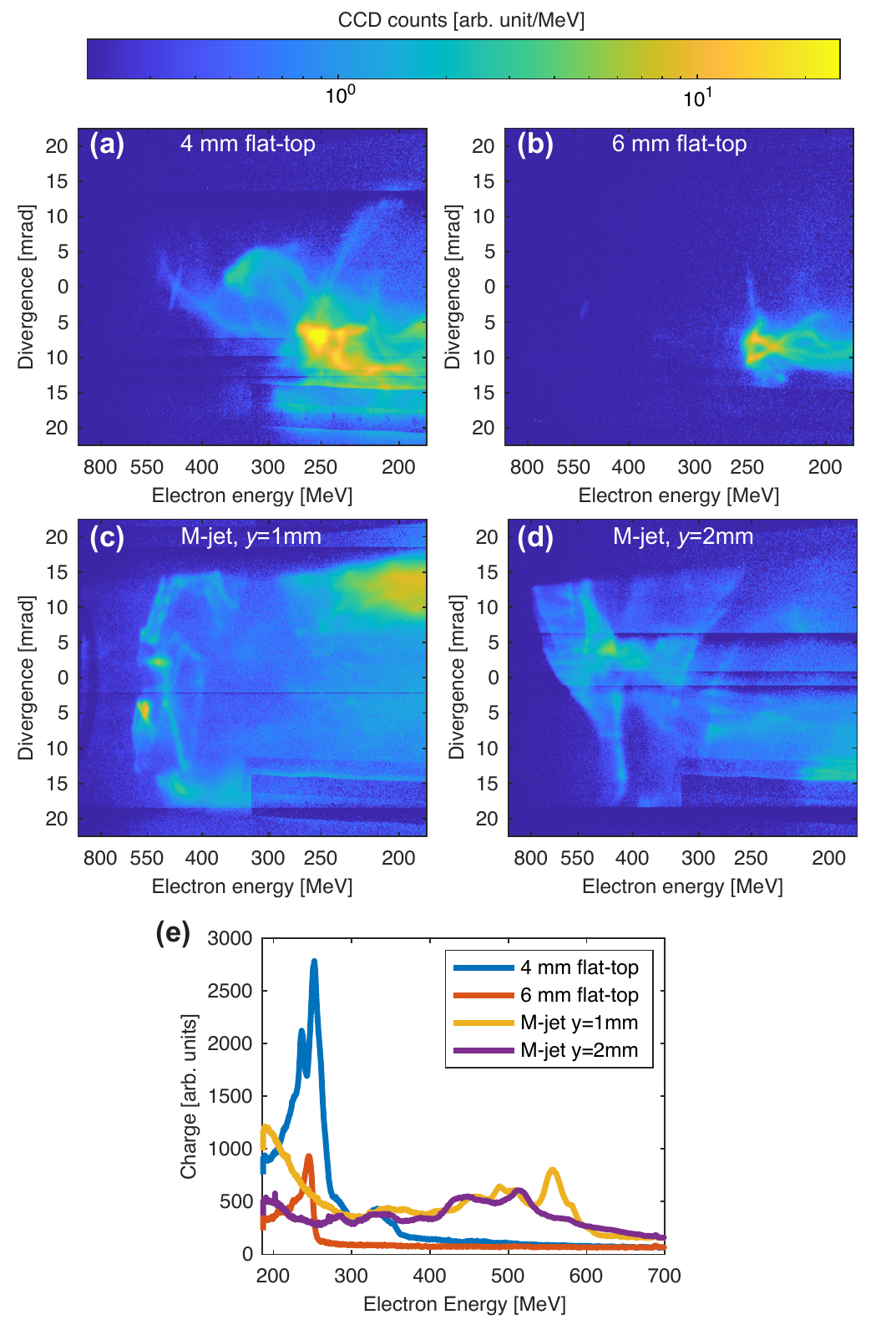}
\caption{Electron Spectra Angularly-resolved in the Laser-Polarization Plane. The electron spectra correspond to the X-ray spectra of Figure \ref{fig2_Xrays}a. The spectrum generated by the 4 mm flat-top gas jet (a) has a large divergence and high charge (see Table \ref{tab:table2}), while the divergence of the 6 mm jet (b) is more typical for LWFAs operated at unmatched plasma densities. The spectrum generated by the M-jet (c) extends to higher energies with a significantly larger divergence. The observed electron spectra, in particular the structured narrow-band, large-divergence features around 570 MeV agree well with the coherent betatron oscillations observed in our particle in-cell (PIC) simulations (see Figure \ref{fig4_simulations}). The divergence of the electron beam is clipped by the 37 mrad angular acceptance of the dipole magnet. The spectrum of the M-jet using an increased interaction distance from the gas nozzle (d) has a similarly large divergence but less charge. For better visibility of the details, the plots are plotted on a logarithmic scale. A linear plot of the angular-integrated spectra is shown in (e). \label{fig3_eSpectra}}
\end{figure}

\begin{table}
\caption{\label{tab:table2} Comparison of the generated electron beam spectra shown in Figure \ref{fig3_eSpectra}. The third column is the charge integrated over an energy range, such that the lowest and highest generated critical photon energies lie in our detection range of 2.7-30 keV.}
\begin{ruledtabular}
\begin{tabular}{c|c|c|c}
Gas jet&Total charge&Charge for generated X- & Divergence\\
& &rays in detection limits& \\
& [CCD counts] & [CCD counts]  &  [mrad]  \\ 
& $\times 10^5$ & $\times 10^5$ &   \\\hline 
\multirow{2}*{\shortstack{4 mm \\ flat-top}} & \multirow{2}*{9.0} & \multirow{2}*{8.5} & \multirow{2}*{24} \\
 & & &  \\ \hline
 \multirow{2}*{\shortstack{6 mm \\ flat-top}} & \multirow{2}*{2.8} & \multirow{2}*{2.6} & \multirow{2}*{10} \\
 & & &  \\ \hline
 \multirow{2}*{\shortstack{M-jet \\ $y$=1 mm}} & \multirow{2}*{7.6} & \multirow{2}*{3.7} & \multirow{2}*{$>$ 37} \\
 & & &  \\ \hline
 \multirow{2}*{\shortstack{M-jet \\ $y$=2 mm}} & \multirow{2}*{4.8} & \multirow{2}*{1.7} & \multirow{2}*{$>$ 37} \\
 & & &  \\ \hline
\end{tabular}
\end{ruledtabular}
\end{table}

We simultaneously measure the spectra of the electron bunches with angular resolution in the plane of the laser polarization (see Figure \ref{fig3_eSpectra}). Although the spectra for all the jets have a broad energy distribution, their charge and angular distributions are significantly different. In particular, for the 6 mm flat-top density profile, we observe the smallest beam divergence, lowest charge and lowest peak energy (see Table \ref{tab:table2}).  In stark contrast, for the 4 mm flat-top and the M-jet we observe a significantly larger divergence, which in case of the M-jet is even clipped by the angular acceptance of our dipole spectrometer of 37 mrad. From the electron beam divergence it becomes clear that only a fraction of the generated photons are observed by our X-ray camera that has an 11 mrad acceptance angle. While the 4 mm flat-top and M-jet both have large divergences, the total charge is significantly larger in case of the 4 mm jet. The maximum electron energy of the M-jet is significantly higher than that of the 4 mm jet, which is significantly higher than that of the 6 mm jet. Both, the 4 mm and the M-jet show energetically narrow features with large divergence. These features indicate off-axis electron injection at a specific position in the gas jet and the performance of coherent betatron oscillations throughout the subsequent density profile until termination of the gas jet. The large divergence, high charge and comparably low beam energy of the 4 mm jet in combination with the comparably weak X-ray emission indicates off-axis electron injection near the end of the gas jet during the density downramp. In case of the M-jet, the lower charge compared to the 4 mm jet, a higher maximum electron energy and stronger X-ray emission suggests off-axis injection during the density downramp of the first jet followed by subsequent acceleration and X-ray emission over a longer distance and possible loss of electrons due to the longitudinal bubble dynamics during the second density peak. Both, the lower maximum energy and lower charge of the 6 mm-long gas jet indicates electron injection and severe laser depletion well before the end of the gas jet, which leads to loss of electrons and decrease in beam energy due to dephasing. The stronger X-ray emission compared to the 4 mm jet despite a lower charge indicates earlier electron injection and betatron emission over a longer distance. These dynamics are qualitatively well in agreement with the dynamics that we observe in our our particle in-cell (PIC) simulations (see section \ref{sec:PICSim}). 

The electron and X-ray parameters of the M-jet can be controlled by the density profile. Specifically, the steepness of the gradients in the modulated density profiles of the M-jets does not significantly impact the divergence and maximum beam energy, but the profile with less steep gradients decreases the accelerated charge, which has also been observed in numerical studies \cite{Ekerfelt:2017}. Due to these specific dynamics for each jet, the measured electron energies and charge after the termination of the jet have only limited correlation with the X-ray emission. Nevertheless, the measured divergence is indicative of the transverse electron momentum at the exit of the jet, which in the bubble regime is mainly determined during the electron injection and largely unchanged over the acceleration distance.

\section{Particle in Cell (PIC) Simulations}
\label{sec:PICSim}
\begin{figure*}
\includegraphics[width=1.8\columnwidth]{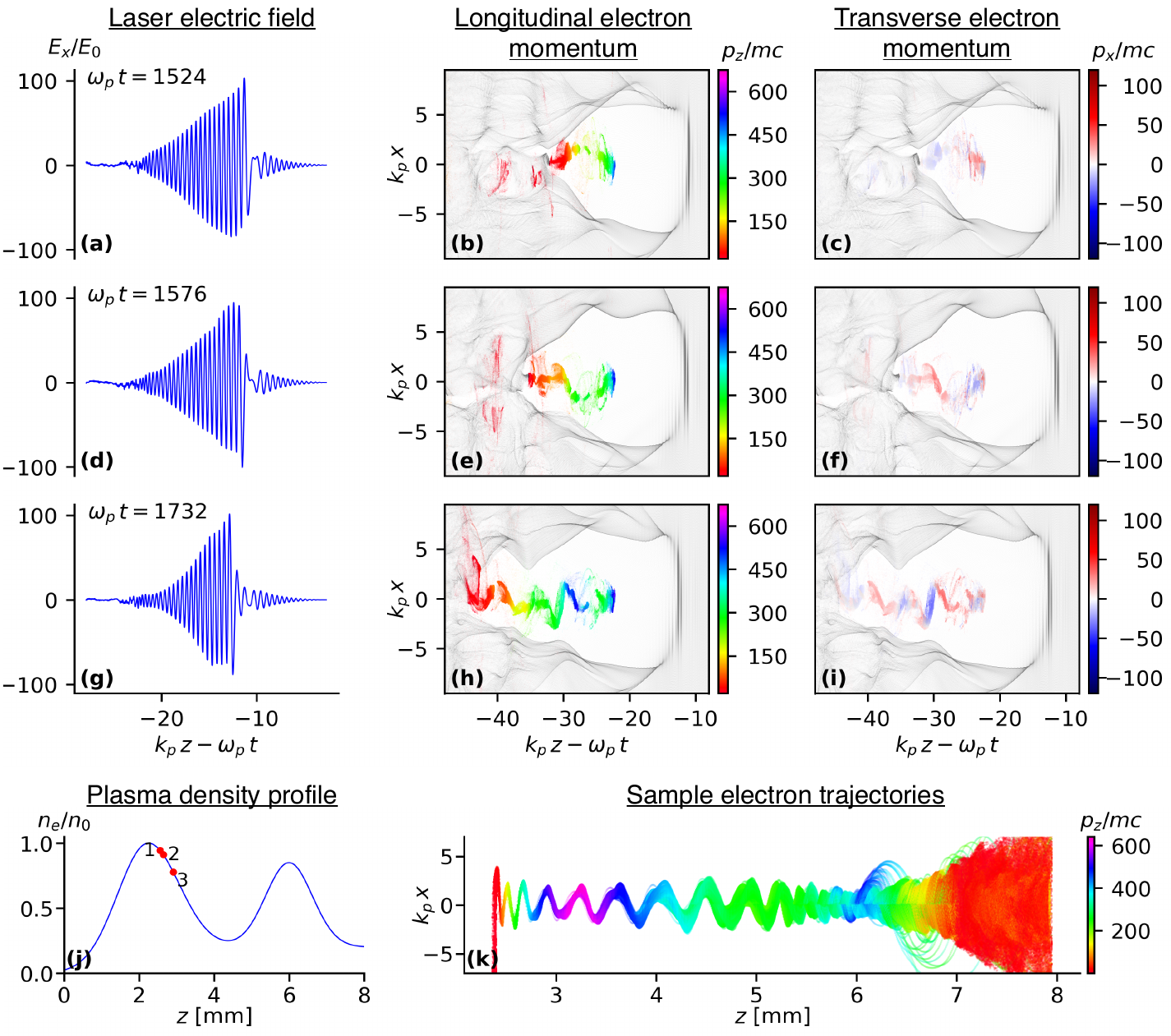}
\caption{Simulated betatron source dynamics (also see movie in SM). The evolution of the laser, bubble and the injected electron beam is shown for different positions along the jet indicated in (j) as ``1" (a)--(c), ``2" (d)--(f) and ``3" (g)--(i). The laser is moving to the right (i.e. in positive $z$-direction). The asymmetric longitudinal laser pulse profile leads to a transverse asymmetry in the bubble shape and off-axis electron injection during the density downramp (a)--(c). The field of the laser front is depleted by a half cycle, leading to a field spike with opposite sign and an asymmetry into the opposite direction (d)--(f). The controlled off-axis injection leads to a correlated transverse momentum distribution (c), (f), and (i) and coherent electron oscillations in the laser polarization ($x,z$) plane. Near the plasma density minimum, the bubble evolves into a super-structure, highly suppressing longitudinal accelerating fields (g)--(i). This leads to electron propagation with nearly constant energy as can be seen from the trajectories of 1429 macro particles from near the head of the electron bunch (k). The laser electric field $E_x$ is normalized to $E_0=mc^2 k_p/e$, the longitudinal and transverse electron momenta (color coded) are normalized to $mc$ and the plasma density to the peak density $n_0 = 1 \times 10^{19} \, \mathrm{cm}^{-3}$. \label{fig4_simulations}}
\end{figure*}

\subsection{M-jet}
We have performed two-dimensional PIC simulations with EPOCH \cite{Arber2015} to better understand the electron dynamics of the betatron source. In the simulations, the gas jet was represented by a combination of two Gaussian profiles that were fitted to the experimentally measured profile as shown in Figure~\ref{fig4_simulations}. The peaks of the jet are  $1\times10^{19}\,\mathrm{cm}^{-3}$ and  $8.5\times10^{18}\,\mathrm{cm}^{-3}$. The simulation domain extended to $\pm 168\,\mu\mathrm{m}$ in the transverse direction with 1487 grid points and used a moving window with $252\,\mu\mathrm{m}$ in the longitudinal direction with 8903 grid points.  The laser pulse, with a FWHM duration of 35 fs, dimensionless vector potential $a_0 = 3.5$ and Rayleigh length $363\,\mu\mathrm{m}$, corresponding to an intensity of $2.62 \times 10^{19} \,\, \mathrm{W/cm}^2$, was launched with a focus located $840\,\mu\mathrm{m}$ from the left hand boundary. A quiet start was used with macro-particles loaded with 16 macro-particles per cell $-67\,\mu\mathrm{m}$ and $67\,\mu\mathrm{m}$ and with 1 macro-particle per cell on the remainder of the domain. The time step was 0.01588/$\omega_p$. The laser is polarized in the ($x$-$z$ plane).

The simulations of the M-jet (movie in Supplementary Material) show a significant laser pulse evolution during the first density peak resulting in complex dynamics of the bubble boundary. This leads to electron injection with specific initial conditions, in particular a large longitudinally-correlated transverse momentum. As a result, the injected electrons perform collective large-amplitude betatron oscillations during the subsequent propagation (Figure \ref{fig4_simulations}). In the laser conditioning section during the first density peak, the laser pulse undergoes substantial self-steepening and pulse compression, which leads to a nearly step-like temporal field distribution of the laser front (Figure \ref{fig4_simulations}a) \cite{Kalmykov2011}. This field distribution causes a transverse asymmetry in the bubble shape (Figure \ref{fig4_simulations}b) \cite{Nerush2009}. The local depletion of the laser front \cite{Decker1996} results in oscillating positive and negative amplitudes of the leading laser-field spike during further propagation. This leads to a bubble asymmetry in the laser polarization direction that changes direction in lock-step with the reversed sign of the leading field amplitude. As a result, the bubble boundary and in particular the large plasma density peak at the back of the bubble perform transverse oscillations in the laser polarization plane (Figure \ref{fig4_simulations}). Due to the longitudinal bubble expansion during the density downramp in the injection section, electrons from this transversely oscillating high-density peak are injected into the accelerating fields at the back of the bubble. Because of the off-axis position, the electrons have a large correlated transverse momentum when injected, which leads to coherent large-amplitude transverse betatron oscillations in the laser polarization plane as they propagate (Figure \ref{fig4_simulations}k)\cite{Nerush2009,Ma2016}. The longitudinally-correlated transverse momentum of the electrons can be seen in Figure \ref{fig4_simulations}i. Despite a large transverse emittance integrated over the whole bunch, the local transverse slice emittance is small. Near the plasma density minimum, the combination of the laser evolution, increased plasma wavelength, lower plasma fields and injected charge (beam loading) leads to an elongated bubble super-structure, extending over multiple buckets (Figure \ref{fig4_simulations}h and i). As a result, the longitudinal bubble fields are highly suppressed and the kinetic energy remains nearly constant for a large fraction of the oscillating electrons over an extended spatial region along the jet. The quasi-constant electron energy and coherent oscillations over a large distance can be seen in example trajectories of 1429 macro particles picked from a region near the front of the electron bunch (Figure \ref{fig4_simulations}k). Due to the lower bubble fields during the density depression at the center of the M-jet, the electron beam experiences a transverse expansion and a small increase in the betatron oscillation amplitude around 3.5 - 5 mm \cite{TaPhuoc2008,Guo2019}. However, this is not the main cause of the large-amplitude oscillations and enhanced betatron emission. During the subsequent density upramp of the second peak, the plasma wavelength decreases and a single-bucket bubble structure is slowly reestablished. As a result, electrons performing betatron oscillations are mainly confined to the first bucket, while electrons further back are lost (see movie in SM). Due to the increase in the bubble phase velocity, only a small fraction of electrons at the head slowly loose energy as they are propagating in a decelerating phase, while a large fraction of electrons are accelerated and gain energies similar to that of the bunch front. This leads to high-brightness X-ray betatron emission from electrons that perform coherent large-amplitude betatron oscillations with almost constant kinetic energy over a long propagation distance in the radiator section. Eventually, the laser pulse is nearly fully depleted and a bubble is driven by the injected electron bunch. The generated radiation for this density profile is significantly enhanced compared to flat-top gas jets. Both, the experimentally observed X-ray and electron beam properties agree well with the simulated kinematics of this process.

\subsection{Jet comparison}
\begin{figure}
\includegraphics[width=1\columnwidth]{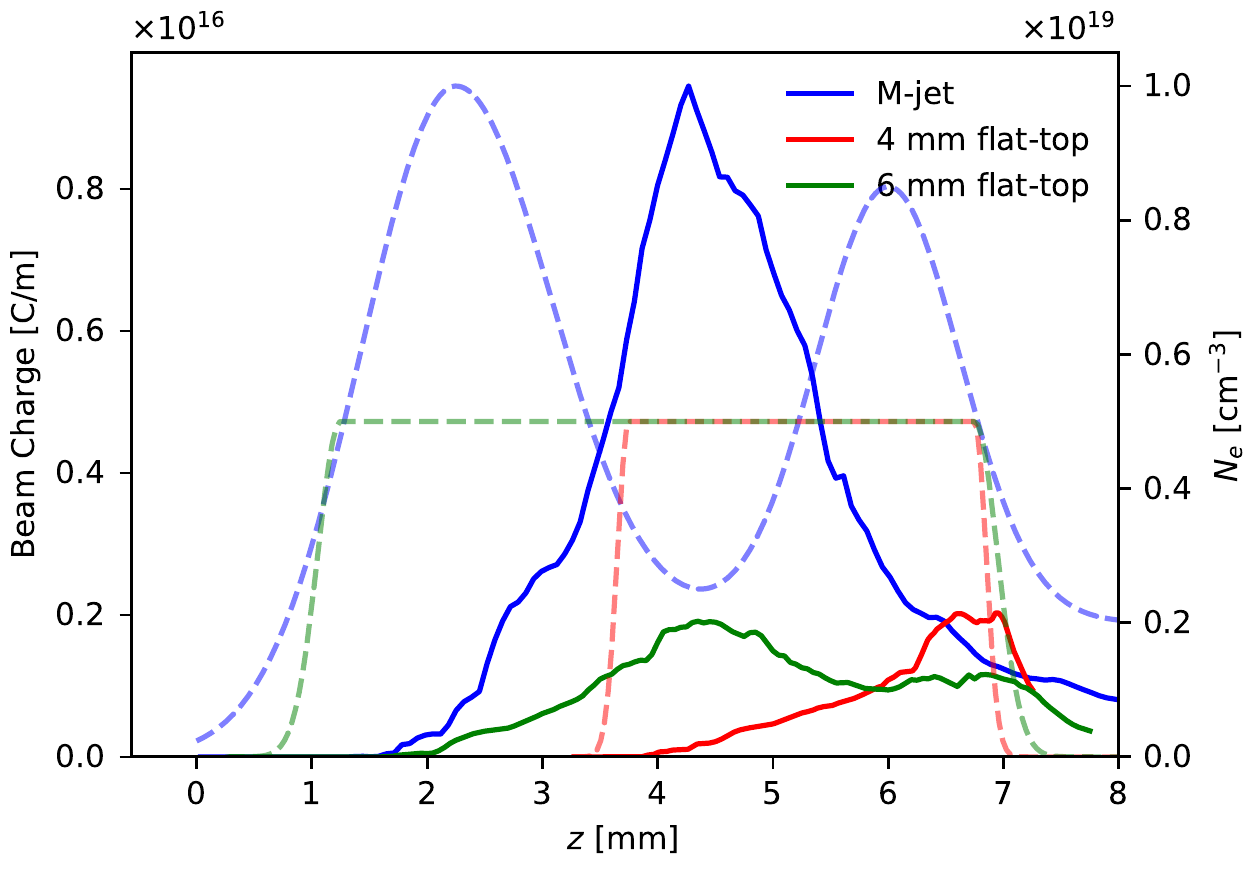}%
\caption{Comparison of the charge for the different jets. The simulation results of the charge near the laser propagation axis ($-15 < k_py < 15$) with a beam energy of $\gamma > 20$ is plotted as a function of the laser propagation distance $z$ (solid line, left axis) for the density profiles (dashed line, right axis) for the M-jet (blue), the 4 mm flat-top (red) and the 6 mm flat-top (green).  \label{fig5_jetcomparison}}
\end{figure}

We have also simulated the comparison between the different jet density profiles. The charge in an area near the laser propagation axis ($-15 < k_py < 15$) and with a beam energy of $\gamma > 20$  is shown as a function of the laser propagation distance (Figure \ref{fig5_jetcomparison}). For the M-jet (blue), the copious amount of charge that is injected during the downramp after the first density peak has a maximum near the density depression and is subsequently slowly lost during the second density peak as the laser gets depleted. For the 6 mm flat-top jet (green), charge is injected after some propagation of the laser. The self-focusing and self-evolution during this propagation leads to an elongation of the bubble that results in further injection. During the subsequent propagation, charge is slowly lost due to laser depletion. In contrast, for the 4mm flat-top jet, charge is mainly injected during the density downramp at the exit of the jet. During the propagation through the jet, the laser self-evolves to a highly asymmetric longitudinal shape similar to that of the M-jet after the first density peak. Correspondingly, this leads to injection with a large transverse momentum.

\section{Discussion}
\label{sec:Discussion}
The measured X-ray and electron spectra indicate different electron beam dynamics in the three jets. The 4-mm flat-top jet generates electron beams with the highest charge and a large divergence while emitting the weakest on-axis X-ray flux. This indicates only a very limited distance during which radiation is generated, which suggests that the electrons are injected near the jet exit at the density downramp. This also agrees with an estimated acceleration distance of approximately 1 mm that is required to reach the observed electron energies for a plasma density of $n_e = 5 \times 10^{18} \,\, \mathrm{cm^{-3}}$. During the propagation through the jet, the laser significantly evolves and the large electron beam divergence suggests that the electrons are injected into a transverse asymmetric bubble driven by such a pulse. After injection, the electrons emit radiation during approximately only 2 betatron oscillations. The reasonable agreement of the X-ray spectrum with the estimated critical energy deduced from the measured electron bunch (see section \ref{sec:X-ray Spectra}) suggest that most of the betatron radiation is emitted from electrons with energies close to their measured spectrum at the exit of the gas jet and that the electrons do not undergo substantial dephasing. 

The electron beams of the 6-mm flat-top jet have the lowest measured charge and smallest divergence while generating a significantly higher X-ray flux than the 4-mm jet particularly at lower photon energies. The measured X-ray spectrum extends well beyond the critical energy of 5 keV, estimated from the measured electron beam parameters. This suggests that the betatron radiation is mostly emitted during the interaction from within the jet. In this case, electrons are continuously injected into the bubble \cite{Kalmykov2011} with a comparably large transverse momentum that is however significantly smaller than in the case of a self-evolved laser. The length of the jet is longer than the dephasing and laser depletion length, which results in a reduction of the electron energy and a loss of charge due to scattering out of the bubble. As the X-rays are generated over a longer distance throughout the electron propagation distance, this leads to the emission of X-ray pulses with a higher flux and higher photons energies than what is expected from the measured electron spectra at the jet exit.

The measured electron beams generated by the M-jet have the largest divergence with a charge that is less than that of the 4 mm-jet, while emitting the highest X-ray flux. The observed electron and X-ray spectra agree with the dynamics described in section \ref{sec:PICSim}, where the laser self-evolves during the first density peak into a highly asymmetric longitudinal shape. During the downramp of the first jet, electrons are transversally injected and X-rays are mostly generated throughout the density depression and the second density peak. Some of the injected charge is likely lost during the interaction. The M-jet combines the features of the other jets, namely the transverse injection of the 4 mm jet with the long emission length of the 6 mm jet for X-ray generation that is enhanced by the large betatron oscillation amplitude and by the density profile that allows X-ray emission over an extended length. While in the experiment it was not possible to measure the effect of only one of the density peaks, the 4 mm jet (at lower density) shows a qualitatively similar behavior to that of only the first density peak. These dynamics are consistent with simulations (see Figures \ref{fig4_simulations} and \ref{fig5_jetcomparison}). 

Increasing the distance between the gas nozzle exit and the laser interaction leads to a change in the plasma density profile, in particular increasing the width of the peaks, decreasing the density slopes and decreasing the peak-to-valley density ratio. In the experiment, this leads to a decrease in both, the injected charge and the on-axis X-ray flux. Simulations have shown that the slope of the density downramp has a strong effect on the electron beam parameters, in particular the amount of injected charge, the transverse emittance and the bunch duration \cite{Ekerfelt:2017}. This demonstrates that by varying the density profile, it is possible to control the X-ray parameters.

\section{Summary and Conclusion}
We experimentally demonstrate the novel Transverse Oscillating Bubble Enhanced Betatron Radiation (TOBER) generation regime that enables significant enhancement and control of laser-driven betatron radiation. In particular, we observe the generation of significantly more X-ray photons compared to standard targets using the same laser parameters. The X-ray source parameters are enhanced and controlled through manipulating the amplitude of the transverse betatron oscillations by using a laser pulse with a highly asymmetric temporal shape that has a nearly step-like rise. This leads to off-axis electron injection and coherent betatron oscillations, similar to those in a permanent-magnet wiggler. The source can be decomposed into three sections: (i) laser pulse shaping, (ii) electron injection and (iii) radiation generation. We demonstrate that this can be realized using a tailored plasma density profile. While longitudinally tailored density profiles have been used to improve the electron beam quality \cite{Geddes2008,Brantov2008,Zhang_2015}, here we use a specifically tailored profile to significantly improve the X-ray beam quality. In our proof-of-principle experiments we use a profile that consists of two peaks separated by a density depression. Here, the laser propagation in the first peak leads to the asymmetric temporal pulse shape through self-evolution. The off-axis electron injection is initiated during the subsequent density downramp. The radiation is mainly generated during the density depression and in the second density peak. We demonstrate that the X-ray pulse properties can be modified by changing the plasma density profile.

The method can be used to readily optimize the X-ray properties for specific applications. The plasma density profile enables control of multiple X-ray properties, including photon energy, number of generated photons and spectral intensity. The method also has the potential to adjust source properties, such as source size, beam divergence and likely pulse duration to the requirements of the specific application. Our concept has the potential for even further enhancement and control of the X-ray parameters and conversion efficiency. We expect that the X-ray beams can be scaled to higher peak brilliance through modifications in the plasma profile, by a higher-power driver laser or by using laser-generated electron bunches as drivers \cite{Ferri2018} to extend distance over which radiation is emitted. Conversely, the plasma profile can also be optimized to generate beams with higher average brilliance using lasers with lower peak power, operating at higher repetition rates. 

The control and enhancement of X-ray parameters is of particular importance for the development of future variable and tunable compact, high-repetition rate X-ray sources that can be readily used in applications. 

\section{Methods}
\subsection{X-ray Spectrum Analysis}
We extract the X-ray spectra from the measured absorption filter transmission using a forward-propagation fitting method. To this end we first process the camera images using a despeckling algorithm to eliminate hot pixels, reduce the measured constant background given by the Pb filter transmission and perform a background gradient normalization using the direct beam in-between filters to account for beam nonuniformities. Experimental measurements for each absorption filter material are extracted by averaging the transmitted signal over the corresponding detector regions taking close to beam center. Standard deviations within these regions, prior to image processing, are used for measurement uncertainty estimates with appropriate scaling to account for image correction. The X-ray spectra are extracted via a Forward-Fitting Monte Carlo (MC) method. As input spectra, we assume a model constrained to a sum of synchrotron spectra given by equation (\ref{eqn:synchrotron_spectra}). We use a sum of synchrotron spectra to account for the broad distribution and dynamics of electron energies, betatron amplitudes and periods, which evolve during the interaction. We found that a sum of two spectra give significantly improved the fit over a single spectrum. Due to a correlation of spectra with different critical energies, we limit the model to a sum of two spectra to avoid over-fitting the data. We iteratively randomly sample the multivariate parameter space (spectral amplitudes $A_i$ and critical energies $E_{\mathrm{crit},i}$) and forward-propagate the corresponding spectra through each filter, including the beamline and camera efficiency. We then compare to the experimental data through a least-square fit and determine the best fit parameters by minimizing the $\chi^2$ value. For each shot we perform the Forward-Fitting MC procedure on 2000 independent data sets. Shots with a $\chi^2$ per degree of freedom value of $\chi^2/N > 0.2$ were excluded from the analysis. We analyze this large parameter space sample to evaluate the goodness of the fit using the bootstrap method \cite{Press2007} and to determine asymmetric point-wise confidence bands. Specifically, for each photon energy value we find the 1-sigma distribution around our best fitting model, which is indicated as confidence bands in Figure \ref{fig2_Xrays}. From the single shot statistics, we calculate the average spectra and confidence bands using a weighted statistical approach. Shot-to-shot statistics are determined by point-wise averaging and appropriate uncertainty propagation considering the five shots with the highest signal intensity. The energies of the fits were limited to a range of 2.7 -  30 keV.

\section{Supplementary Material}
The transmission through the filters used in the spectrometer and a movie of the particle-in cell (PIC) simulation can be seen in the Supplementary Material.

\subsection*{Data Availability Statement}
The data that support the findings of this study are available from the corresponding author upon reasonable request.
 
\begin{acknowledgments}
This material is based upon work supported by the Air Force Office of Scientific Research (AFOSR) under award number FA9550-15-1-0125. TK and BAS where supported but the US DoE under award number DE-SC0018363 and by NFS under award number PHY-1535678. The experiment was conducted at the Extreme Light Laboratory at the University of Nebraska-Lincoln. Part of this work was completed utilizing the Holland Computing Center of the University of Nebraska, which receives support from the Nebraska Research Initiative. Part of this research was performed using resources provided by the Open Science Grid, which is supported by the National Science Foundation award \#2030508.
\end{acknowledgments}


\providecommand{\noopsort}[1]{}\providecommand{\singleletter}[1]{#1}%

\end{document}